\documentclass[aps, prl, twocolumn, superscriptaddress, groupedaddress]{revtex4}
\pdfoutput=1

\usepackage{amsmath}
\usepackage{graphics}

\begin{document}

\title{Weak and strong interactions between dark solitons and
  dispersive waves}

\author{I. Oreshnikov}
\affiliation{Department of Physics, St. Petersburg State University, 198504, Ulyanovskaya st. 1, Peterhof, St. Petersburg, Russian Federation}
\author{R. Driben}
\affiliation{ITMO University 197101, Kronverksky pr. 49, St. Petersburg, Russian Federation}
\affiliation{Department of Physics and CeOPP, University of Paderborn, Warburger Str. 100, D-33098 Paderborn, Germany}
\author{A. V. Yulin}
\affiliation{ITMO University 197101, Kronverksky pr. 49, St. Petersburg, Russian Federation}


\date{\today}

\begin{abstract}
The effect of mutual interaction between dark solitons and dispersive waves is investigated numerically and analytically. The condition of the resonant scattering of dispersive waves on dark solitons is derived and compared against the results of numerical simulations. It is shown that the interaction with intense dispersive waves affects the dynamics of the soltons strongly changing their frequencies and accelerating or decelerating the solitons. It is also demonstrated that two dark solitons can form a cavity for dispersive weaves bouncing between the two dark solitons. The differences of the resonant scattering of the dispersive waves on the dark and bright solitons are discussed. In particular we demonstrate that two dark solitons and dispersive wave bouncing in between them create solitonic cavity with convex ``mirrors'' unlike the concave ``mirror'' in case of the bright solitons.
\end{abstract}

\maketitle

\section{Introduction}

Optical dark solitons were predicted \cite{Hasegawa} and experimentally realized \cite{Experiment1} several decades ago and since then attracted considerable attention \cite{Obzor, Agrawal, Zhao, Boris, Dumitru, Bang, PT} in optical community. The interest to the dark solitons is motivated by their fascinating fundamental properties and
by possible applications for telecommunications \cite{Obzor, Agrawal}. Dark solitons exist in the normal dispersion regime and they are characterized by the dip on top of the flat intensity background. Vectorial dark solitons \cite{Turitsyn} as well as coupled bright and dark solitons were considered \cite{Afanasyev, Trillo}. Also the dynamics of dark solitons in presence of higher order dispersive effects was well covered \cite{Menyuk, Mahalingam, Milian, Obzor, Agrawal}. Several studies have also addressed emission of dispersive waves (DWs) by dark solitons \cite{Menyuk, Milian} as well as emission of radiation by shock waves \cite{Mateo2} propagating in normal dispersion region.

Discussing the interaction of dispersive waves and bright solitons it is relevant to remind that extensive studies were performed on scattering of DWs on bright solitons, with soliton’s properties remaining invariant \cite{yulin, pre, Efimov, Skryabinoverview, Conforti}. In particular it was also demonstrated that strong resonant scattering of DWs on solitons can accelerate or decelerate the solitons with consequent frequency up- or downshifts \cite{DribenMitschke, Demircan}. This mechanism can be viewed as a method of all-optical switching \cite{Demircan, Demircan2, Tartara} or alternative technique for generation of broad and coherent supercontinuum \cite{Demircan2}. A very interesting scenario occurs when DWs emitted by one soliton interacts with a neighboring solitons through the dispersive waves trapped between the solitons. The multiple scattering of DW on solitons results in the mutual attraction of the solitons and eventually to their collision \cite{Resonator}.  Furthermore, these collisions can cause fusion of two solitons \cite{Fusion}. This phenomenon was proven to be responsible for the appearance of multiple soliton knot patterns \cite{DribenMitschke, Resonator2, NC} during the complex supercontinuum generation process. Very recently the effect of strong variations of the solitons trajectories under the action of the DW  was experimentally demonstrated \cite{Tartara}. An experimental observation of the solitonic cavities confining DW was very recently reported in \cite{Kudlinski}.

This work aims to address the interaction of dark solitons with external DWs and to explore the possibility of manipulation of dark solitons by DWs.
We will also demonstrate a possibility to create solitonic cavities made from dark solitons and DWs oscillating in between them.

\section{Interaction of weak dispersive waves with dark solitons}

For a weak DW interacting with the bright soliton it is possible to construct an analytic theory predicting the location of the resonance frequencies \cite{yulin, pre}. Here we develop such a theoretical approach based on the perturbation theory for the case of dark solitons interacting with weak DWs.

Dynamics of the dimensionless amplitude $u(z, t)$ in normal dispersion regime with significant third-order dispersion can be described by a generalized nonlinear Schr\"odinger equation
\begin{equation}
  \label{eq:NLSE}
  i \partial_{z} u
    - \frac{1}{2} \partial_{t}^{2} u
    - \frac{i}{6} \beta_{3} \partial_{t}^{3} u
    + |u|^{2} u = 0
\end{equation}
We seek for a solution in the form of a dark soliton perturbed by a sum of incident DW and scattered radiation, generated by interaction between DW and the soliton
\begin{equation*}
  u(z, t) = \left[
      \psi_{sol}(z, t)
    + \psi_{inc}(z, t)
    + \psi_{sc}(z, t)
  \right] e^{i u_{0}^{2} z}
\end{equation*}
where $\psi_{sol} = u_{0} \tanh(u_{0} t)$. Assuming that both DW and scattered radiation are small compared to the soliton background $u_{0}$ and retaining only the linear terms we can write the following equations for $\psi_{inc}$ and $\psi_{sc}$
\begin{align}
  \label{eq:IncEquation}
  i \partial_{z} \psi_{inc}
  - \frac{1}{2} \partial^{2}_{t} \psi_{inc}
  - \frac{i}{6} \beta_{3} \partial^{3}_{t} \psi_{inc}
  + u_{0}^{2} \left(
    \psi_{inc} +  \psi_{inc}^{*}
  \right) = 0 \\
  \label{eq:ScEquation}
  i \partial_{z} \psi_{sc}
  - \frac{1}{2} \partial^{2}_{t} \psi_{sc}
  - \frac{i}{6} \beta_{3} \partial^{3}_{t} \psi_{sc}
  + (2 \psi_{sol}^{2} - u_{0}^{2}) \psi_{sc}
  + \psi_{sol}^{2} \psi_{sc}^{*}
  \nonumber \\
  = \frac{i}{6} \beta_{3} \partial_{t}^{3} \psi_{sol}
  - 2 (\psi_{s}^{2} - u_{0}^{2}) \psi_{inc}
  -   (\psi_{s}^{2} - u_{0}^{2}) \psi_{inc}^{*}
\end{align}
Right hand's side of equation includes three driving terms: first one generates Cherenkov radiation of the soliton (which has been already analyzed in \cite{Menyuk}), while the other two correspond to FWM process between DW and the soliton. Seeking for a plane wave solution for \eqref{eq:IncEquation} by setting
\begin{equation*}
  \psi_{inc} = \psi_{+} e^{  i k(\omega_{inc}) z - i \omega_{inc} t}
         + \psi_{-} e^{- i k(\omega_{inc}) z + i \omega_{inc} t}
\end{equation*}
we arrive at the dispersion relation
\begin{equation}
  \label{eq:DispersiveRelation}
  k(\omega_{inc}) =
    \frac{1}{6} \beta_{3} \omega_{inc}^{3}
    \pm \omega_{inc} \sqrt{\frac{1}{4} \omega_{inc}^{2} + u_{0}^{2}}
\end{equation}
The condition of the resonance excitation of the delocalized eigenmodes of the medium can be obtained by the analysis of the asymptotic of the eigenfunctions when where $\psi_{sol}^{2}$ approaches $u_{0}^{2}$. The equation then turns into a version of \eqref{eq:IncEquation} with additional forcing terms and for the FWM resonance frequencies $\omega_{sc}$ we can write
\begin{equation}
  \label{eq:ResonanceFrequencies}
  \frac{1}{6} \beta_{3} \omega_{sc}^{3}
    \pm \omega_{sc} \sqrt{\frac{1}{4} \omega_{sc}^{2} + u_{0}^{2}}
  = \frac{1}{6} \beta_{3} \omega_{inc}^{3}
    \pm \omega_{inc} \sqrt{\frac{1}{4} \omega_{inc}^{2} + u_{0}^{2}}
\end{equation}

\begin{figure}[h]
  \centering
  \includegraphics{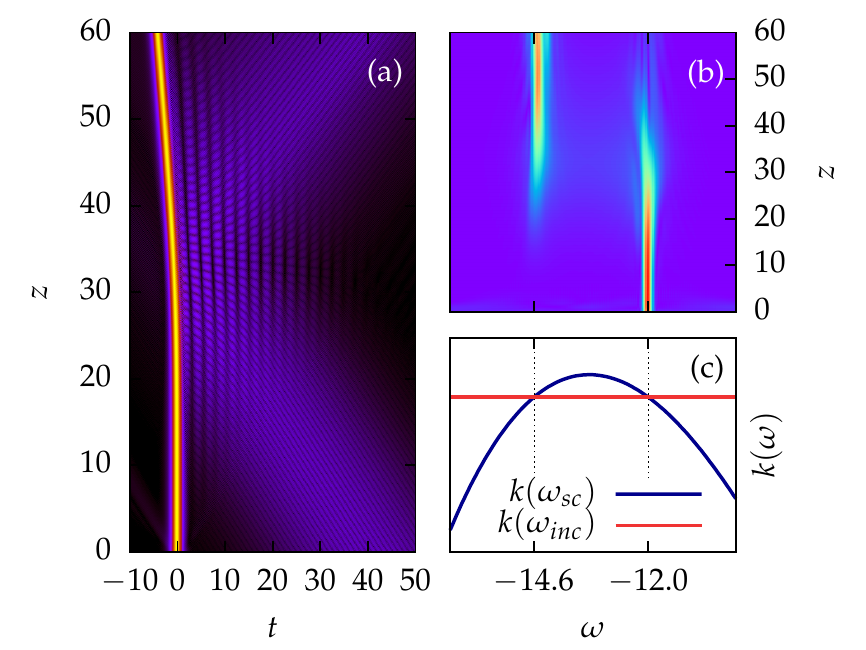}
  \caption{A relatively weak DW ($A = 0.10$) scattering on a dark soliton. (a) is the field intensity minus the background $\left||u(z, t)|^{2} - u_{0}^{2}\right|$, (b) is the spectral density $|u(z, \omega)|$ in the relevant region of frequency domain, (c) displays a graphical solution to \eqref{eq:ResonanceFrequencies}, where the blue curve and the red line are, respectively, the left and the right hand's sides of the equation.}
  \label{fig:DarkSolitonScattering}
\end{figure}

\begin{figure}[h]
  \centering
  \includegraphics{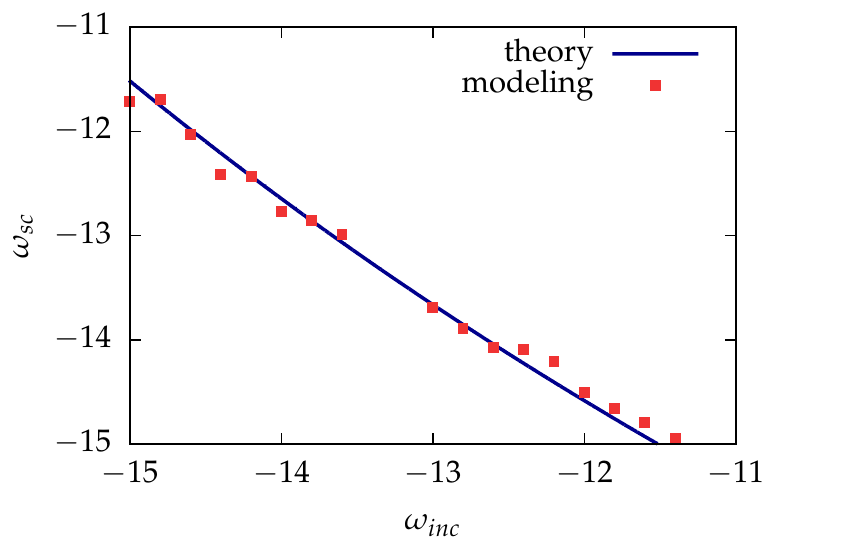}
  \caption{Scattered frequency $\omega_{sc}$ as a function of incident frequency $\omega_{inc}$. Solid line is the theoretical prediction according to \eqref{eq:ResonanceFrequencies}, markers are found by numerical modeling.}
  \label{fig:ScatteredVsIncident}
\end{figure}

To confirm the analysis above we have conducted a series of numerical simulations. \eqref{eq:NLSE} was solved using split-step Fourier method with the initial condition
\begin{equation*}
  u(0, t) =
    u_{0} \tanh(u_{0} t)
    + A \exp(-(t - t_{1})^{2} / \tau^{2})
      \exp(-i \omega_{inc} t)
\end{equation*}
where $A$ is the amplitude, $t_{1}$ is the initial position of the center and $\tau$ is the width of DW. We started with a relatively weak dispersive wave with amplitude of $A = 0.1$ reflecting off the soliton with $u_{0} = 1$. A typical result of simulation is presented on Fig. \ref{fig:DarkSolitonScattering}. As it is shown on Fig. \ref{fig:DarkSolitonScattering}(b) around the collision point ($z \approx 35$) the frequency  of the dispersive radiation changes. The new frequency of the DW  is very well predicted by the resonance condition \eqref{eq:ResonanceFrequencies}, see Fig. \ref{fig:DarkSolitonScattering}(c) showing the graphic solution of the resonance condition. To further test \eqref{eq:ResonanceFrequencies} we have varied the  frequency $\omega_{inc}$ of the incident pulse of DW and measuring the scattering frequency $\omega_{sc}$ at the end of simulation distance. The numerically measured frequencies are shown in  Fig. \ref{fig:ScatteredVsIncident} by red squares alongside with the prediction of the resonance condition  shown by the black curve. As one can see, the resonance condition predicts the frequency of the scattered DW $\omega_{sc}$ very well.

It should be noted here that the resonance condition predicts two resonant frequencies even if the dispersion charactersitics of the DW has only one branch. In our case the dispersion characteristics of the small excitation on the background of the dark soliton has two branches and the resonant scattering between the DW belonging to different branches of the the dispersion characteristics is also possible. However only one of the resonant scattering channels is efficient and in this paper we discuss only this kind of scattering.

\section{Interaction of strong dispersive waves with dark solitons and dark solitonic cavities}

\begin{figure}[t]
  \centering
  \includegraphics{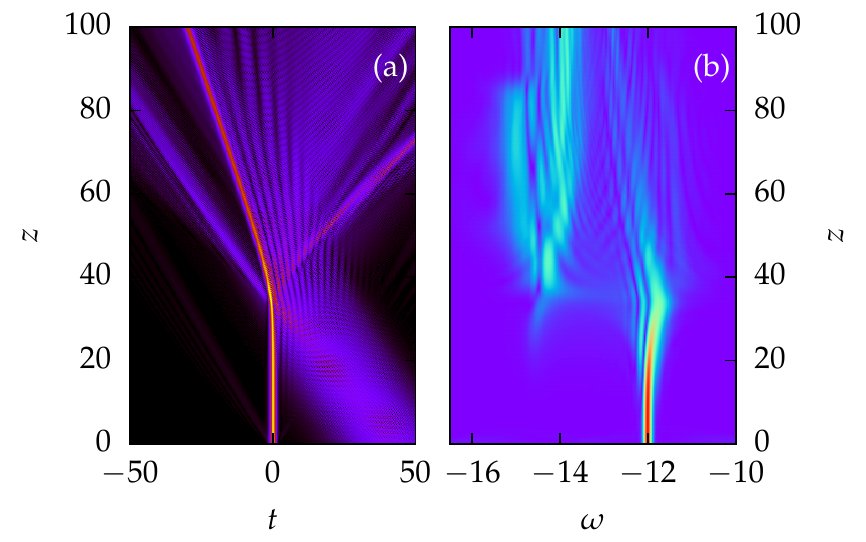}
  \caption{An intensive DW ($A = 0.20$) scattering on a dark soliton. (a) is the field intensity minus the background $\left||u(z, t)|^{2} - u_{0}^{2}\right|$, (b) is the spectral density $|u(z, \omega)|$ in the relevant region of frequency domain.}
  \label{fig:DarkSolitonScatteringStrongDispwave}
\end{figure}

The interaction of an intensive DW with a soliton is a more complicated case with richer dynamics, which, unfortunately, cannot be adequately described by the analytics developed in the previous section. The main drawback of the simple linearized theory is that we have assumed that the soliton does not change while interacting with DW, which is, strictly speaking, not entirely true even for a weak DW. Collision with more intensive DWs seems to have a substantial effect on the solitons and, as a consequence, on the spectral structure of the DW itself. For example, twofold increase in the amplitude of DW ($A = 0.20$) leads to DW pushing the soliton from its original trajectory as it is demonstrated on Fig. \ref{fig:DarkSolitonScatteringStrongDispwave}.
Importantly, the steering of the trajectory in case of the dark soliton takes place in the direction opposite to that of the incident DW.
This is completely different from the case of the bright solitons that is steered toward the incident DW.
Looking at the process in the frequency domain (see Fig. \ref{fig:DarkSolitonScatteringStrongDispwave}(b)) one can notice that scattered wave is considerably
broadened spectrally. This can be explain by the fact that the soliton frequency is changing under the action of the incident DW and therefore different parts of the
incident pulse scatter into different frequencies enriching the spectrum of the DW.

\begin{figure}[t]
  \centering
  \includegraphics{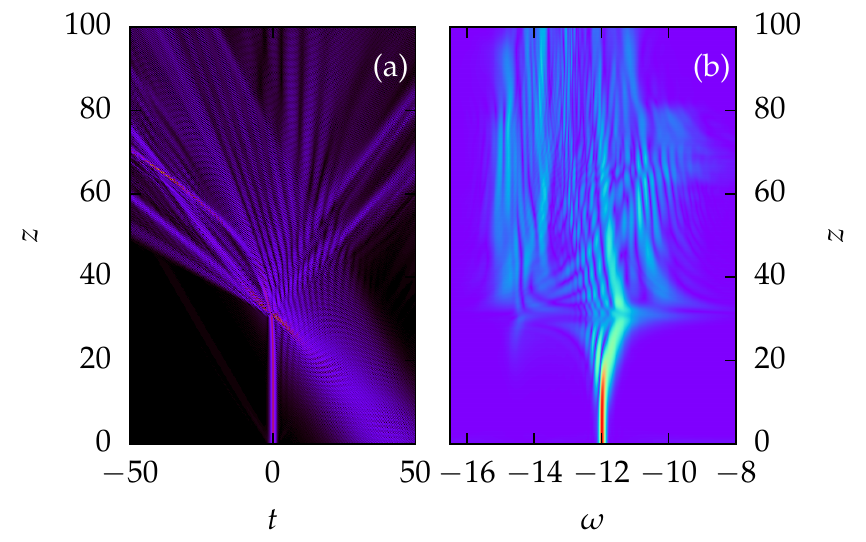}
  \caption{A very intensive DW ($A = 0.30$) colliding with a dark soliton triggering its collapse. (a) is the field intensity minus the background $\left||u(z, t)|^{2} - u_{0}^{2}\right|$, (b) is the spectral density $|u(z, \omega)|$ in the relevant region of frequency domain.}
  \label{fig:DarkSolitonDisintegration}
\end{figure}

Further increase of the intensity of DW leads to a decomposition of the black soliton that manifest itself by a frequency continium in the spectral domain.
Fig. \ref{fig:DarkSolitonDisintegration} demonstrates an outcome of the simulation in which an intensive DW with the amplitude $A = 0.30$ collides with the
soliton causing disintegration of the soliton.

\begin{figure}[t]
  \centering
  \includegraphics{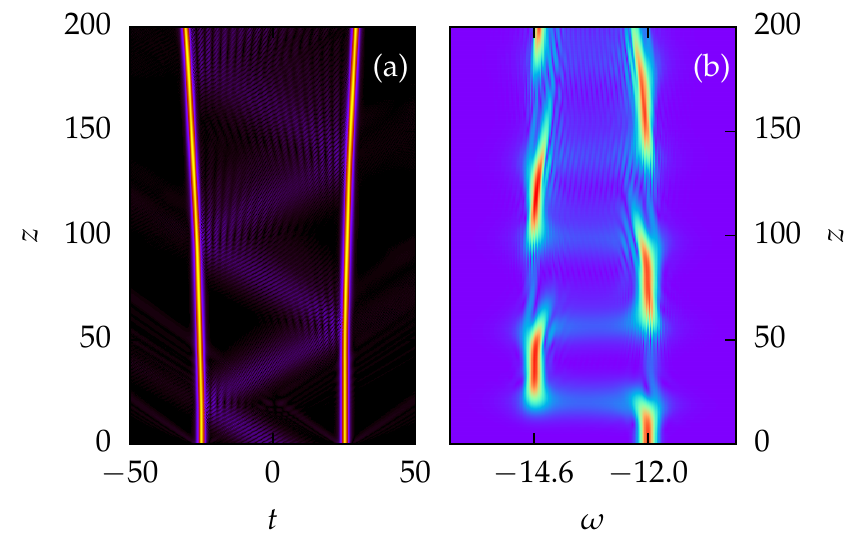}
  \caption{A very weak DW ($A = 0.05$) trapped in a two soliton
    cavity. (a) is the field intensity minus the background
    $\left||u(z, t)|^{2} - u_{0}^{2}\right|$, (b) is the spectral
    density $|u(z, \omega)|$ in the relevant region of frequency
    domain.}
  \label{fig:DarkSolitonCavityWeakDispwave}
\end{figure}

\begin{figure}[t]
  \centering
  \includegraphics{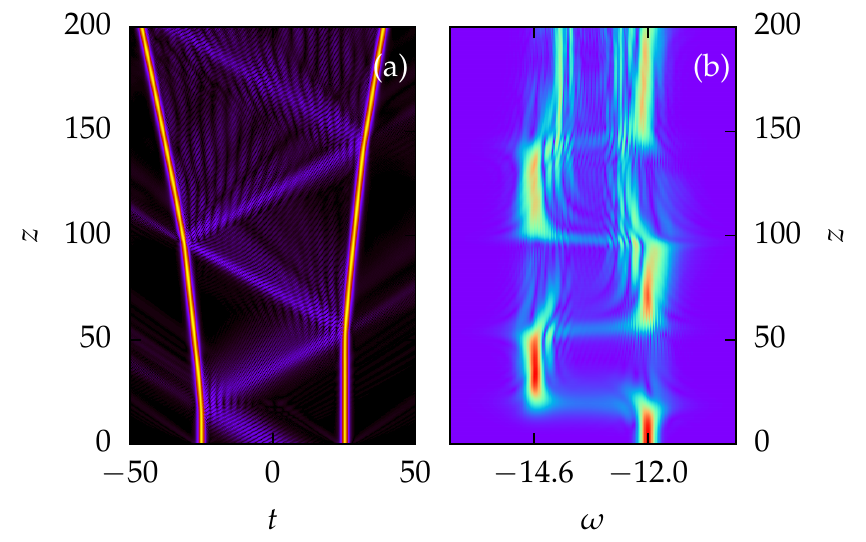}
  \caption{A weak DW ($A = 0.1$) trapped in a two soliton cavity. (a)
    is the field intensity minus the background $\left||u(z, t)|^{2} -
      u_{0}^{2}\right|$, (b) is the spectral density $|u(z, \omega)|$
    in the relevant region of frequency domain.}
  \label{fig:DarkSolitonCavity}
\end{figure}

A very interesting configuration arises when we launch a relatively
weak DW in a free space bound by two dark solitons. Initial condition
in this case is changed to
\begin{align*}
  u(0, t) &= u_{cvt}(t)
           +  A \exp(- t^{2} / \tau^{2})
                \exp(-i \omega_{inc} t) \\
  u_{cvt}(t) &= \begin{cases}
    - u_{0} \tanh(u_{0} (t + t_{1})), & \quad t < 0 \\
    + u_{0} \tanh(u_{0} (t - t_{1})), & \quad t \ge 0
  \end{cases}
\end{align*}
where $t_{1}$ is the soliton launch position. In this case two
solitons act as boundaries trapping DW inside a soliton cavity. DW
periodically bounces off the walls of the cavity while changing its
carrier frequency in a periodical fashion, as it is illustrated in
Figs.~\ref{fig:DarkSolitonCavityWeakDispwave},~\ref{fig:DarkSolitonCavity}.
Since we are dealing with multiple scatterings, soliton trajectory can
be significantly perturbed even by a relatively weak DW with amplitude
$A = 0.1$ (see Fig.~\ref{fig:DarkSolitonCavity}). There is a major
difference from the case of two bright solitons creating a solitonic
cavity \cite{Resonator, Resonator2, Kudlinski}. The two dark solitons
create a ``convex mirror'' like cavity when reflecting the DW and not
``concave mirror'' cavity as produced by the bright solitons.

\section{Conclusions}
The resonant scattering of weak and strong dispersive waves on dark solitons is studied analytically and numerically. An anlytical formula predicting the frequency of the scattered wave was derived for the case of weak dispersive waves. Numerical simulations show that the resonant condition predicts the frequency of the scattered waves with very high accuracy. The interaction between the solitons and relatively intense DW was studied by means of pure numerical simulations. It is demonstrated that strong dispersive waves can affect dark soliton's trajectory strongly which opens a way to control dark solitons by dispersive waves. An important observation is that the bright soliton trajectories bends against the direction of the incident DW and the dark solitons trajectories bend in the direction of the incident DW. It was also demonstrated that two dark solitons allow creation of a solitonic cavity being able to trap dispersive waves making them  bouncing between the solitons behaving like a couple of convex mirrors. The radiation trapped between the solions results in the dispersive-waves mediated repulsion of the solitons at the distances much larger than the characteristic size of the solitons.

\section*{Funding Information}

R.D and A.V.Y gratefully acknowledges the support by the Russian Federation Grant 074-U01 through ITMO Early Career Fellowship scheme.

\end{document}